\documentclass[prl,twocolumn,nofootinbib,final]{revtex4}
\usepackage[utf8]{inputenc}
\usepackage{graphicx,color}
   \usepackage{amsmath,amssymb,pifont}
      \usepackage{simplewick}

\newcommand{\nn}{\nonumber}

\newcommand{\Tr}{\mathrm{Tr}}

\renewcommand{\(}{\left(}
\renewcommand{\)}{\right)}

\renewcommand{\vec}[1]{\mathbf{#1}}

\begin{document}
\title{Self-contained definition of the Collins-Soper kernel}
\author{Alexey A. Vladimirov}
\email{Alexey.Vladimirov@physik.uni-regensburg.de}
\affiliation{Institut f\"ur Theoretische Physik, Universit\"at Regensburg,\\
D-93040 Regensburg, Germany}

\begin{abstract}
The rapidity anomalous dimension (RAD), or Collins-Soper kernel, defines the scaling properties of transverse momentum dependent distributions and can be extracted from the experimental data. I derive a self-contained nonperturbative definition that represents RAD without reference to a particular process. This definition makes possible exploration of the properties of RAD by theoretical methods on one side, and the properties of QCD vacuum with collider measurements on another side. To demonstrate these possibilities, I compute the power correction to RAD, its large-b asymptotic, and compare these estimations with recent phenomenological extractions.
\end{abstract}
\maketitle

\textbf{Introduction.} The nontrivial structure of the QCD vacuum raises a lot of fundamental and yet unsolved problems, such as mechanisms of quark confinement and hadronization. As a matter of fact, there is a little number of experimental observables that test properties of the QCD vacuum. In this Letter, I demonstrate that the evolution kernel of transverse momentum dependent (TMD) distributions is exclusively sensitive to the structure of QCD vacuum and thus is a valuable tool to study it.

The rapidity anomalous dimension (RAD), or Collins-Soper kernel, was introduced in Refs.\cite{Collins:1981uk,Collins:1981va}, as a part of the factorized formula, which accumulates the double-logarithm contributions. In later works, it has been shown that RAD is universal for different processes and receives nonperturbative (NP) corrections. The rigorous formulation of the TMD factorization theorem \cite{Collins:2011zzd,Becher:2010tm,GarciaEchevarria:2011rb,Vladimirov:2017ksc} has identified RAD as an independent NP function that contains the information about soft-gluon exchanges between partons and dictates the evolution properties of TMD distributions. Until recently, the NP terms of TMD evolutions were not examined individually, but as a constituent of the resummation exponent or TMD distributions. Although it does not necessarily contradict the theory, it makes it difficult to split effects related to different sides of strong dynamics. One of the main messages of this Letter is that RAD is an important function with a rich physical background, and thus must be seen as an independent distribution. 

Despite the long history of RAD, very little is known about its NP nature from the theory side. Apart from a general identification that the NP part exists, I know only a few works that are dedicated to this problem at least partially \cite{Tafat:2001in,Lee:2006nr,Schweitzer:2012hh,Becher:2013iya,Collins:2014jpa,Scimemi:2016ffw}. In this Letter, I would like to draw attention to this gap in the theory. As an initial step, I provide a field-theoretical and model-independent definition of RAD detached from the cross section formula. Given the definition, RAD can be used as a self-contained phenomenological function of QCD, which measures properties of QCD vacuum. To demonstrate the power of the derived definition, I compute the leading terms of operator product expansion (OPE) and compute RAD within a simplistic model.

\textbf{Appearance of RAD.} The cross section for the Drell-Yan pair production at small transverse momentum is described by the TMD factorization formula
\begin{eqnarray}\label{xsec1}
\frac{d\sigma}{dq_T}&=&\sigma_0(Q/\mu)\int d^2 b e^{-i \vec b\cdot \vec q_T}
\\\nn&& \times \(\frac{Q^2}{\zeta}\)^{-2\mathcal{D}(b,\mu)}F_1(x_1,b;\mu,\zeta)F_2(x_2,b;\mu,\zeta),
\end{eqnarray}
where $Q$ is the virtuality of a photon, and $q_T$ its transverse momentum\footnote{Scalar products of traverse vectors in bold font have are defined as $(\vec b \vec q_T)=-(bq_T)$. Consequently, $\vec b^2=-b^2>0$.}. The functions $F$ are TMD distributions, and $\mathcal{D}$ is RAD\footnote{There is no common notation for RAD. The other popular notations are $-\tilde K/2$ \cite{Collins:1981uk,Collins:2011zzd,Aybat:2011zv}, $-F_{q\bar q}$ \cite{Becher:2010tm}.}. Similar formulas describe the small-transverse momentum regime of semi-inclusive deep inelastic scattering (SIDIS), and $e^+e^-\to h_1h_2+X$-process. In Eq.(\ref{xsec1}) and in the following, I omit the flavor indices for brevity, keeping in mind that RAD depends on the color representation of the parton. 

A distinctive feature of TMD distributions is their dependence on two scales\cite{Aybat:2011zv,Scimemi:2018xaf}: the factorization scale $\mu$ and the scale of rapidities separation $\zeta$. The dependence on these scales is given by a pair of equations, where the first one is an ordinary renomalization group equation for the scale $\mu$ and the second one is
\begin{eqnarray}\label{dF/dzeta=D}
\frac{d F(x,b;\mu,\zeta)}{d\ln \zeta}=-\mathcal{D}(b,\mu)F(x,b;\mu,\zeta).
\end{eqnarray}
The integrability condition for the pair of evolution equations gives the dependence of RAD on $\mu$,
\begin{eqnarray}\label{RAD:RGE}
\frac{d\mathcal{D}(b,\mu)}{d\ln \mu}=\Gamma_{\text{cusp}}(\mu),
\end{eqnarray}
where $\Gamma_{\text{cusp}}$ is the anomalous dimension for the cusp of lightlike Wilson lines. In a conformal field theory, RAD is equivalent to the soft anomalous dimension and entirely perturbative \cite{Vladimirov:2016dll,Vladimirov:2017ksc}. In QCD RAD is a general NP function, although it still inherits some properties of an anomalous dimension, such as additive structure of renormalization group equation, see Eq.(\ref{RAD:RGE}).

\begin{figure}[t]
\includegraphics{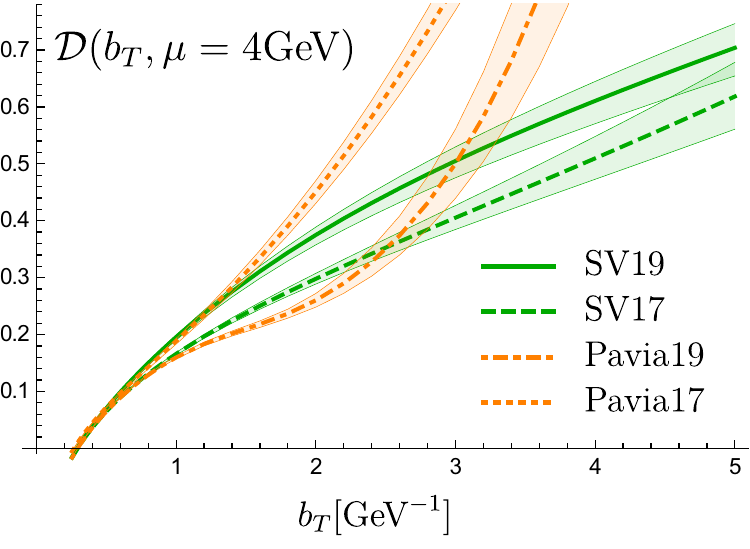}
\caption{\label{fig:phenomenology} Comparison of extracted values of RAD. The lines labeled as SV19, SV17, Pavia19 and Pavia17 correspond to  Refs.\cite{Scimemi:2019cmh},\cite{Scimemi:2017etj},\cite{Bacchetta:2019sam}, and \cite{Bacchetta:2017gcc}.}
\end{figure}

Equation (\ref{dF/dzeta=D}) essentially mixes the definitions of two NP functions: a TMD distribution and RAD. For that reason, the separation of these functions with the data is a nontrivial phenomenological task. Nonetheless, it could be done observing that RAD governs the $Q$ behavior of the cross section, whereas $F$'s govern the $x$ behavior. Therefore, analyzing a global set of data with a large span in $x$ and $Q$, it is possible to decorrelate these functions. Such global studies were made  recently \cite{Bacchetta:2017gcc,Scimemi:2017etj,Bertone:2019nxa,Scimemi:2019cmh,Bacchetta:2019sam}. The values of RAD obtained in these works are shown in Fig.\ref{fig:phenomenology}. Clearly, there is no agreement between these extractions for $b>2$GeV$^{-1}$. Another observation is that extraction based on the joined data of Drell-Yan and SIDIS cross sections \cite{Scimemi:2019cmh,Bacchetta:2017gcc} provide a higher value of RAD at $b\sim 1$GeV$^{-1}$ in comparison to extraction based only on the Drell-Yan data \cite{Scimemi:2017etj,Bacchetta:2019sam}. These contradictions could be resolved by adding more low-$q_T$ data in the analysis, or by some alternative approaches to access RAD. One of promising approaches is the recently proposed methods to compute RAD with lattice QCD \cite{Ebert:2018gzl,Ji:2019sxk,Vladimirov:2020ofp}.

\textbf{Definition of RAD.} To derive the self-contained expression for RAD, I take a step backward in the derivation of Eq.(\ref{xsec1}) and recall the origin of scale $\zeta$. At an intermediate stage, the expression for the cross section has the form $d\sigma\sim \tilde F_1\times S\times \tilde F_2$ \cite{Collins:2011zzd,GarciaEchevarria:2011rb}, where $\tilde F$ are unsubtracted TMD distributions, and $S$ is the TMD soft factor. Each of these terms contains the rapidity divergence(s) that cancel in the product. To obtain (\ref{xsec1}), the soft factor is factorized into parts with only rapidity divergences related to a particular lightlike direction. Afterwards, they are combined with $\tilde F$ into physical TMD distributions \cite{Echevarria:2012js,Vladimirov:2017ksc,Chiu:2012ir}. The scale $\zeta$ in the definition of a physical TMD distribution (\ref{dF/dzeta=D}) is the scale of rapidity divergence factorization. Thus, the soft factor is the primary object to define RAD.

\begin{figure}[t]
\includegraphics[width=0.47\textwidth,trim={0 0 0 2cm},clip]{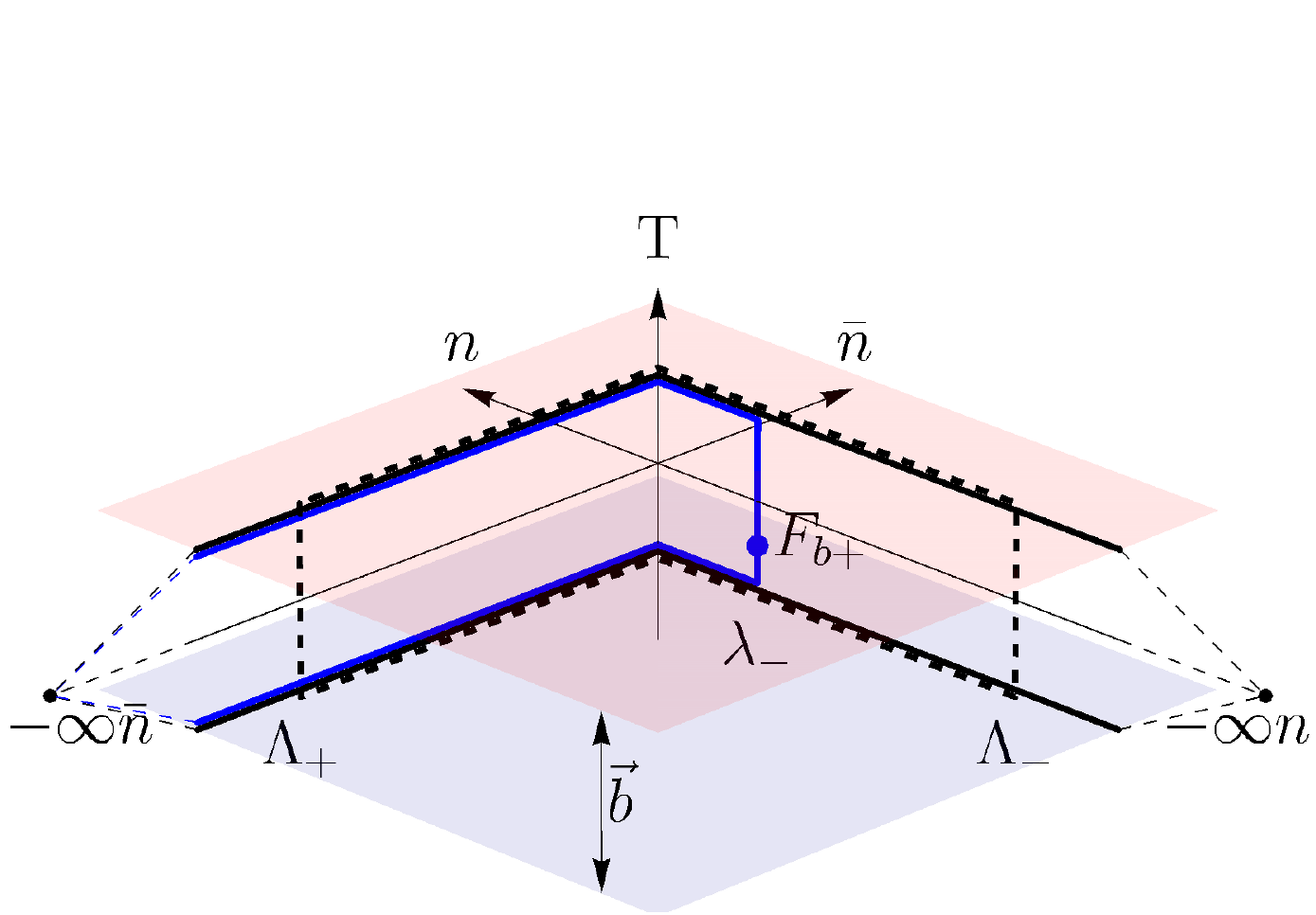}
\caption{\label{fig:contour} Contours defining the TMD soft factor (in the Drell-Yan kinematics) and its derivatives. Axes $n$ and $\bar n$ are lightlike ($n^2=\bar n^2=0$), and the axis T is transverse. The black (blue) solid line shows contour $C$ ($C'$). The black dashed lines show the contour $C_\Lambda$. The blue dot shows the insertion of gluon strength tensor.}
\end{figure}

The TMD soft factor  is defined as
\begin{eqnarray}
S_C(b,\mu)=\frac{\Tr}{N_c}\langle 0|W_C|0\rangle Z^2_{S}(\mu), 
\end{eqnarray}
where $W_C=P\exp(ig\int_Cdx^\mu A_\mu(x))$ is a gauge link along the contour $C$ (see fig.\ref{fig:contour}), $Z_{S}$ is the renormalization factor for lightlike cusps. In Ref.\cite{Vladimirov:2017ksc} it has been proven that the TMD soft factor with a properly designed regularization has the general form
\begin{eqnarray}\label{S=exp(D+b)}
S_C(b,\mu)=\exp\(2\mathcal{D}(b,\mu)\ln(\varrho)+B(b,\mu)+ ...\),
\end{eqnarray}
where $\varrho$ is the Lorenz-invariant combination of parameters of rapidity divergence regularization($\varrho\to 0$). The function $B$ is the finite part of the soft factor, and the dots denote terms vanishing at $\varrho\to 0$. Consequently, RAD can be obtained from the TMD soft factor as
\begin{eqnarray}\label{D=dS/dr}
\mathcal{D}(b,\mu)=\frac{1}{2}\lim_{\varrho \to 0} \frac{d\ln S_C(b,\mu)}{d \ln \varrho}.
\end{eqnarray}

The expression (\ref{S=exp(D+b)}) is a general one, but it is difficult to use outside of the perturbation theory. The main complication is the definition of an appropriate rapidity divergence regulator. To guarantee (Eq.\ref{S=exp(D+b)}) and make use of Eq.(\ref{D=dS/dr}), the regulator must be given on the level of the operator, preserve the gauge invariance, and fully regularize rapidity divergences without generation of extra infrared divergences. None of the commonly used regulators in perturbative calculations regulators (see e.g.Refs.\cite{Collins:2011zzd,Becher:2010tm,GarciaEchevarria:2011rb,Echevarria:2015byo,Li:2016axz,Li:2016ctv}) fulfill these requirements entirely. The discussion of the drawbacks in common regularizations can be found in Refs.\cite{Vladimirov:2017ksc,Echevarria:2015byo,Echevarria:2016scs}. All these requirements can be fulfilled by a deformation of the contour $C$ such that it does not touch lightlike infinities \cite{Vladimirov:2017ksc}. The most straightforward deformation is the contour $C_\Lambda$ shown in Fig.\ref{fig:contour}. In this case, the parameters $\Lambda_\pm$ regularize rapidity divergences at both infinities and $\varrho=(\Lambda_+\Lambda_-)^{-1}$. 

The regularized soft factor $S_{C_\Lambda}$ is a function of $\varrho$ and $b^2$ (and $\mu^2$), because these are the only nonzero scalar products in the task. The regularization is removed by limits $\Lambda_+\to\infty$ and $\Lambda_-\to\infty$, but since the dependence on $\Lambda$'s is given by a single variable $\varrho$, one of these limits is obsolete. For definiteness, I fix $\Lambda_-=\lambda_-$. The derivative with respect to $\varrho=(\Lambda_+\lambda_-)^{-1}$ can be replaced by derivative over $\lambda_-$, and Eq.(\ref{D=dS/dr}) turns into
\begin{eqnarray}\label{D=dS/dl}
\mathcal{D}(b,\mu)=\frac{1}{2}\lim_{\Lambda_+ \to \infty} \frac{d\ln S_{C_\Lambda}(b,\mu)}{d \ln \lambda_-}.
\end{eqnarray}
The action of the derivative is
\begin{eqnarray}\label{RAD-00}
&& \mathcal{D}(b,\mu)=Z_{\mathcal{D}}(\mu)+
\\\nn &&~ \lim_{\Lambda_+ \to \infty}\lambda_-\frac{ ig}{2}\frac{\Tr \int_0^1 d\beta \langle 0|F_{b+}(-\lambda_-n+ b \beta)W_{C_\Lambda}|0\rangle}{\Tr \langle 0|W_{C_\Lambda}|0\rangle},
\end{eqnarray}
where $F_{b+}(x)=b^\mu n^\nu F_{\mu\nu}(x)$, with $F_{\mu\nu}$ being a gluon-field strength tensor, and $Z_{\mathcal{D}}(\mu)=d \ln Z_{S}/d\ln \lambda_-$. The contour in the numerator starts and ends at the point $(-\lambda_-n+b \beta)$, so the numerator is the Wilson loop with insertion of the gluon strength tensor. The limit $\Lambda_+\to\infty$ turns the contour $C_\Lambda$ (with finite $\lambda_-$) to the contour $C'$ shown in fig.\ref{fig:contour} in blue, and
\begin{eqnarray}\label{RAD-main}
&& \mathcal{D}(b,\mu)=
\\\nn && \lambda_-\frac{ ig}{2}\frac{\Tr \int_0^1 d\beta \langle 0|F_{b+}(-\lambda_-n+ b \beta)W_{C'}|0\rangle}{\Tr \langle 0|W_{C'}|0\rangle}+Z_{\mathcal{D}}(\mu).
\end{eqnarray}
Here, the numerator and the denominator have rapidity divergences, which cancel each other. So, to use (\ref{RAD-main}) beyond tree order, a convenient regularization for these divergences should be introduced. The term $Z_{\mathcal{D}}(\mu)=d \ln Z_{S}/d\ln \lambda_-$ removes the ultraviolet divergences. Peculiarly, it is additive rather than multiplicative, which produces the renormalization group equation of the form (\ref{RAD:RGE}), with
\begin{eqnarray}
\frac{dZ_{\mathcal{D}}(\mu)}{d\ln \mu}=\Gamma_{\text{cusp}}(\mu).
\end{eqnarray}
The additional cusps present in $C'$ do not introduces divergences since $(n b)=0$ \cite{Korchemsky:1992xv}.

Despite the the left-hand-side of Eq.(\ref{RAD-main}) has an explicit entry of $\lambda_-$, the expression is independent on it. It is an outcome of the dependence of $S_{C_\Lambda}$ on $\varrho$. Alternatively, the $\lambda_-$ independence can be seen as a consequence of the boost invariance. Different values of $\lambda_-$ can be related by a boost in the $n$ direction. Therefore, $\lambda_-$ occurs in the numerator and the denominator of Eq.(\ref{RAD-main}) only due to the rapidity divergences and cancel in the ratio. The independence on $\lambda_-$ also demonstrates the universality of RAD for Drell-Yan and SIDIS processes, which is dictated by the sign of $\lambda_-$ in the current context.

The expression (\ref{RAD-main}) is the main result of this Letter. In contrast to previous works, the definition (\ref{RAD-main}) gives a direct access to RAD. In the next paragraphs, I demonstrate possible applications of it and make elementary checks.

\textbf{OPE and perturbative computation.} RAD is very well studied in the perturbation theory, where it has been derived up to next-to-next-to-leading order (NNLO) \cite{Vladimirov:2016dll,Li:2016ctv}. All previous calculations have been done by evaluation of the TMD soft factor \cite{Echevarria:2015byo,Li:2016ctv}, or TMD distributions \cite{Echevarria:2016scs,Aybat:2011zv}, with successive identification of rapidity divergent terms. Using Eq.(\ref{RAD-main}) RAD can be computed directly. 

The perturbative calculation is made in the regime $b\ll\Lambda_{QCD}^{-1}$. In this regime, RAD can be written as
\begin{eqnarray}
\mathcal{D}(b,\mu)=\mathcal{D}_0(b,\mu)+\vec b^2 \mathcal{D}_2(b)+(\vec b^2)^2 \mathcal{D}_4(b)+...~,
\end{eqnarray}
where dots designate terms accompanied by a higher power of $\vec b^2$. Each $\mathcal{D}_n$ depends on $b$ only logarithmically, via $\ln(b\mu)$. Importantly, the definition (\ref{RAD-main}) is made for a finite $b$. The limit $b\to 0$ does not exist due to the presence of divergent renormalization constant $Z_{\mathcal{D}}$ that is independent on $b$. Indeed, already at LO $\mathcal{D}_0\sim \alpha_s(\mu)\ln(b\mu).$  The terms with $n>0$ do not depend on $\mu$ explicitly, as it follows from the independence of $Z_\mathcal{D}$ on $b$.

The computation of $\mathcal{D}_n$ can be done, for example, by the background field method, similarly to calculations made in Refs.\cite{Balitsky:1987bk,Scimemi:2019gge}. It is convenient to use the background field in the Schwinger gauge with a reference point at the origin. With this choice, Wilson lines of background gluons turn to unities at $b\to0$, which crucially simplifies the calculation.

The LO contribution to term $\mathcal{D}_0$ is given by a one-loop diagram,
\begin{eqnarray}
\lambda_-\frac{g^2}{2}\int_0^1 d\beta\int_{C'}dx^\mu ~\frac{\Tr}{N_c}\!
\contraction[4pt]{}{F_{b+}}{(-\lambda_-n+ b\beta)}{ A_\mu}
~F_{b+}(-\lambda_-n+ b\beta) A_\mu(x).
\end{eqnarray}
The result of computation in the dimensional regularization ($d=4-2\epsilon$ with $\epsilon>0$, and $\overline{\text{MS}}$-scheme) reads
\begin{eqnarray}
\mathcal{D}_0(b,\mu)=- 2C_Fa_s\Big[\Gamma(-\epsilon)\(\frac{\vec b^2\mu^2}{4 e^{-\gamma_E}}\)^\epsilon\!\!+\frac{1}{\epsilon}\Big]\!+\!\mathcal{O}(a_s^2),
\end{eqnarray}
where $a_s=g^2/(4\pi)^2$, and the value of $Z_S$ is taken from Ref.\cite{Echevarria:2016scs}. This expression coincides with the one derived in \cite{Echevarria:2015byo} at arbitrary $\epsilon$, and in the limit $\epsilon \to 0$ reproduces the well-known result \cite{Collins:1981uk,Collins:2011zzd,Becher:2010tm,GarciaEchevarria:2011rb}
\begin{eqnarray}
\mathcal{D}_0(b,\mu)=2C_Fa_s(\mu)\ln\(\frac{\vec b^2 \mu^2}{4e^{-2\gamma_E}}\)+\mathcal{O}(a_s^2).
\end{eqnarray}
Note, that there is no dependence on $\lambda_-$, as expected.

\begin{figure}[t]
\includegraphics[width=0.25\textwidth]{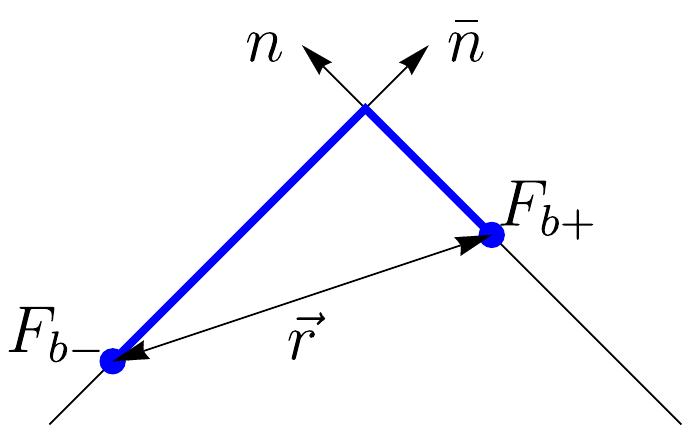}
\caption{\label{fig:op_phi} Structure of the operator that describes the leading power correction to RAD. Blue lines are the gauge links, and dots are insertions of gluon strength tensors.}
\end{figure}

In contrast to $\mathcal{D}_0$, power suppressed terms have tree-order contributions, which are the coefficients of the Taylor series at $b=0$. Each term introduces a new NP function which are matrix elements of gluon strength tensors connected by Wilson lines to the origin. The LO contribution to $\mathcal{D}_2$ is given by
\begin{eqnarray}\label{D2:op}
-g^2\lambda_-\int_0^1 d\beta \!\int_{-\infty}^0 \!\!\!d\sigma \langle 0| F_{b+}(-\lambda_- n)[..]F_{b-}(\sigma \bar n)|0\rangle,
\end{eqnarray}
where $F_{b-}=b^\mu n^\nu F_{\mu\nu}$, and $[..]$ stays for a gauge link (in the adjoint representation) between $F$'s and the origin (see Fig.\ref{fig:op_phi}). It is a particular case of the following matrix element
\begin{eqnarray}
\Phi_{\mu\nu}(x,y)=g^2 x^\alpha y^\beta\langle 0| F_{\mu \alpha}(x)[x,0][0,y]F_{\nu \beta}(y)|0\rangle.
\end{eqnarray}
The matrix element $\Phi_{\mu\nu}$ satisfies $x^\mu\Phi_{\mu\nu}=y^\nu\Phi_{\mu\nu}=0$ and thus can be parametrized by two independent functions
\begin{eqnarray}\label{Phi:param}
&&\Phi_{\mu\nu}(x,y)=\(g_{\mu\nu}-\frac{y_\mu x_\nu}{(xy)}\)\varphi_1(r^2,x^2,y^2)
\\ \nn&&\quad  + \frac{(x_\mu (xy) -y_\mu x^2)(y_\nu (xy) -x_\nu y^2)}{(xy)((xy)^2-x^2y^2)}\varphi_2(r^2,x^2,y^2),
\end{eqnarray}
where $r^2=(x-y)^2$. At  $x^2=y^2=0$, $\varphi_2$ vanishes and only $\varphi_1$ contributes to Eq.(\ref{D2:op}), but at higher orders of perturbative series both terms are present. 

Using the parametrization (\ref{Phi:param}) I receive
\begin{eqnarray}\label{D2:expression}
\mathcal{D}_2(b)=\frac{1}{2}\int_0^\infty d\vec r^2 \frac{\varphi_1(\vec r^2,0,0)}{\vec r^2}+\mathcal{O}(a_s).
\end{eqnarray}
For the first time, the power correction to RAD is expressed in a model-independent way in terms of QCD vacuum correlations. It could be compared to expressions derived in Refs.\cite{Lee:2006nr,Becher:2013iya}, which involves the jet-algorithm-modified QCD vacuum.

The function $\varphi_1$ is unknown, nonetheless, its value could be estimated. In particular, at $\vec r^2\to 0$ it is 
\begin{eqnarray}
\lim_{\vec r^2\to 0} \frac{\varphi_1(\vec r^2,x^2,y^2)}{\vec r^2}=\frac{\pi^2}{36}G_2,
\end{eqnarray}
where $G_2=(g^2/4\pi^2)\langle 0|:F_{\mu\nu}^aF_{\mu\nu}^a:|0\rangle$ is the gluon condensate \cite{Shifman:1978bx}. At large $\vec r^2$, $\varphi_1$ decays at least as $\vec r^{-2}$ (more realistically, it decays exponentially). Assuming $\varphi_1$ has an effective radius $\sim \Lambda^{-1}_{\text{QCD}}$, $\mathcal{D}_2$ can be estimated as
\begin{eqnarray}\label{c2-estimation}
\mathcal{D}_2\sim \frac{\pi^2}{72}\frac{G_2}{\Lambda^2_{\text{QCD}}} \simeq (1.-5.) \times 10^{-2}\text{GeV}^2.
\end{eqnarray}
The values of parameters are taken from the review in Ref.\cite{Narison:2018dcr}. The estimation (\ref{c2-estimation}) is notably a small number. Nonetheless it is in agreement with recent extractions that are collected in the following table (see also Fig.\ref{fig:phenomenology}).
\begin{center}
\begin{tabular}{c||c|c|c|c}
Ref. & \cite{Bacchetta:2017gcc} & \cite{Scimemi:2019cmh}& \cite{Scimemi:2017etj} & \cite{Bacchetta:2019sam} \\
\hline
$\mathcal{D}_2 \times 10^2$GeV$^2$ & $2.8\pm0.5$ & $2.9\pm 0.6$ & $0.7^{+1.2}_{-0.7}$  & $0.9\pm 0.2$
\end{tabular}
\end{center}
These values are obtained with LO approximation at $\mu=2$GeV.  Let me note that earlier considerations, which are often used in high-energy phenomenology, such as BLNY-fit \cite{Landry:2002ix,Su:2014wpa}, use significantly higher values, $\mathcal{D}_2\sim 0.2-0.35$GeV$^2$.

\textbf{Example of NP modeling: stochastic vacuum model.} One of the most promising applications of the expression (\ref{RAD-main}) is the computation of RAD with various NP models. It would help to select appropriate phenomenological ansatz and give an intuitive interpretation for RAD. As an example, I evaluated $\mathcal{D}$ in the stochastic vacuum model (SVM) \cite{DiGiacomo:2000irz}. Although this model cannot be considered realistic, it catches some global features of QCD, such as the area law.

In SVM one assumes that the QCD dynamics is dominated by two-point correlators, whereas multipoint correlators give a negligible contribution. Additionally, one ignores the gauge links connecting fields, assuming their unimportance at large distances. In this way, all gluonic observables are written in terms of two functions $\Delta$ and $\Delta_1$, defined as \cite{DiGiacomo:2000irz}
\begin{eqnarray}\label{def:Delta}
&&g^2 \langle 0|F_{\mu\nu}(x)F_{\alpha\beta}(0)|0\rangle =
\\\nn
&& (g_{\mu\alpha}g_{\nu \beta}-g_{\mu\beta}g_{\nu \alpha})(\Delta(x^2)+\Delta_1(x^2))
 +(g_{\mu\alpha}x_\nu x_\beta
\\\nn 
&& \qquad
-g_{\nu\alpha}x_\mu x_\beta -g_{\mu\beta}x_\nu x_\alpha+g_{\nu\beta}x_\mu x_\alpha)\frac{\partial \Delta_1(x^2)}{\partial x^2}.
\end{eqnarray}
Applying the non-Abelian Stockes theorem to Eq.(\ref{RAD-main}), dropping multipoint correlators, and using Eq.(\ref{def:Delta}), after some simplification I arrive at the expression
\begin{eqnarray}
&&\mathcal{D}(b)=\vec b^2\int_{\vec b^2}^\infty d\vec y^2\(\Delta(\vec y^2)+\frac{\Delta_1(\vec y^2)}{2}\)
\\\nn &&\qquad+\int_0^{\vec b^2}d \vec y^2\Big[\frac{\vec y^2}{2}\Delta_1(\vec y^2)+(2\sqrt{\vec b^2 \vec y^2}-\vec y^2)\Delta(\vec y^2)\Big].
\end{eqnarray}
This expression predicts the linear behavior of $\mathcal{D}$ at large $b$,
\begin{eqnarray}\label{SVM:large-b}
\lim_{\vec b^2\to \infty} \mathcal{D}(b)= \sqrt{\vec b^2}\int_0^\infty d\vec y^2 ~2\sqrt{\vec y^2}\Delta(\vec y^2).
\end{eqnarray}
A similar calculation but in the momentum space has been done in Ref.\cite{Tafat:2001in}. Although the final results could not be compared, some intermediate steps and Eq.(\ref{SVM:large-b}) are in agreement. The integral can be roughly estimated using lattice computations \cite{Bali:1997aj,Meggiolaro:1998yn,Simonov:2018cbk} as $c_\infty\simeq 0.01 - 0.4$GeV. The value can be compared to $c_\infty=0.06\pm 0.01$GeV extracted in Fef.\cite{Scimemi:2019cmh}, which uses the model for $\mathcal{D}$ with the linear asymptotic.

Considering various relations derived in SVM (in particular, the static interquark potential \cite{Brambilla:1996aq}), I found that for the internal consistency of the model one has to demand
\begin{eqnarray}\label{RAD:large-b}
\lim_{\vec b^2 \to \infty} \mathcal{D}(b)\sim (\vec b^2)^{1/2-\delta},\qquad \delta \geqslant 0.
\end{eqnarray}
It significantly restricts the shape of $\mathcal{D}$ at large $b$. In fact, the expression (\ref{RAD:large-b}) disregards almost all models for RAD used in phenomenology, since the dominant part of studies (e.g., Refs.\cite{Bacchetta:2017gcc,Scimemi:2017etj,Landry:2002ix,Su:2014wpa}) use quadratic asymptotic $\mathcal{D}\sim \vec b^2$ or even stronger \cite{Bacchetta:2019sam}. The same or equivalent conclusion as Eq.(\ref{RAD:large-b}) has been also made in Refs.\cite{Tafat:2001in,Collins:2014jpa}.

\textbf{Conclusion.} The expression (\ref{RAD-main}) is the main result of this Letter. This expression is unique in several aspects. It is a definition of a scaling kernel through the matrix elements. It grants the opportunity to study RAD without referring to TMD distributions. It gives a connection between the vacuum structure and the particle scattering. Each of these aspects is a promising direction for further studies.

To provide an elementary check and the demonstration of definition (\ref{RAD-main}) I present the LO perturbative computation at $b\to0$ and recover the well-known expression. It is clear that for perturbative computations, which are already performed at the NNLO level \cite{Vladimirov:2017ksc,Li:2016ctv}, the new definition is not more advantageous than an ordinary one. The main power of the new definition is that it gives a direct operator definition of RAD. It allows us to compute power corrections and apply the nonperturbative modeling for RAD, which is not possible in a standard approach.


The derived LO power correction (\ref{D2:expression}) is model independent. To my best knowledge, it is the first derivation of this object. The expression could be systematically improved by computing higher-order terms. The LO computation predicts a small size of the power correction, which is in agreement with the most recent extractions. The model calculation, performed in SVM, put a serious restriction on the shape of RAD at large values of $b$ (\ref{RAD:large-b}). Altogether, these findings severely constrain the evolution properties of TMD distributions and should be accounted for in the analysis. The calculations are done for RAD of quark TMD distributions. It could be easily repeated for the gluon case. The only modification is the color representation for gauge links. Consequently, all expressions derived in the Letter are also valid for gluon RAD after the Casimir rescaling (that is valid up to N$^3$LO), which consists of the multiplication by $C_A/C_F(=9/4)$. 

The possibility to investigate the QCD vacuum in high energy collisions sounds contradictory to the intuitive picture that the structure of accelerated particles is cleared from low-energy effects. Indeed, the partons do not interact with each other within a highly energetic hadron. Nonetheless, their temperate transverse motion is sensitive to the structure of the underlying vacuum. Therefore, measuring the low-$q_T$ behavior of high-energy scattering at different energies, one examines the QCD vacuum. In fact, the measurements by LHC restricts RAD significantly, as it is shown in Refs.\cite{Bertone:2019nxa,Bacchetta:2019sam,Hautmann:2020cyp}.

\acknowledgements

Author is thankful to V.~Braun, O.Teryaev, A.Sch\"afer, and I.Scimemi for stimulating discussions. This work was supported by DFG (FOR 2926 ``Next Generation pQCD for Hadron Structure: Preparing for the EIC'', project number 430824754).

\bibliography{TMD_ref}
\end{document}